\documentclass[fleqn,usenatbib]{mnras} 

\usepackage[T1]{fontenc}
\usepackage{ae,aecompl}

\usepackage{mathptmx}

\usepackage[dvips]{graphicx}
\usepackage{amsmath}                 
\usepackage{amssymb}                 
\usepackage{bm}                      

\def\bea{\begin{eqnarray}} \def\eea{\end{eqnarray}}
\def\be{\begin{equation}} \def\ee{\end{equation}}
\def\bal#1\eal{\begin{align}#1\end{align}}
\def\bse#1\ese{\begin{subequations}#1\end{subequations}}
\def\pv{\bm{p}}
\def\ra{\rightarrow}
\def\al{\alpha}
\def\eps{\varepsilon}

\def\ms{\,M_\odot}
\def\mmax{M_\text{max}}
\def\fm3{$\text{fm}^{-3}$}

\def\mfmrm{$\rm{\;MeV\,{fm}^{-3}}$\ }
\def\mqm{M_\text{QM}}
\def\rqm{\rho_\text{QM}}
\def\mdu{M_\text{DU}}
\def\xdu{x_\text{DU}}
\def\rdu{\rho_\text{DU}}
\def\r1s0{\rho_{1S0}}
\def\m1s0{M_{1S0}}
\def\bds{B_\text{DS}}

\title
[Cooling of hybrid neutron stars with microscopic equations of state]
{Cooling of hybrid neutron stars with microscopic equations of state}

\author
[J.-B. Wei, G. F. Burgio, H.-J. Schulze, and D. Zappal\`a]
{J.-B. Wei, G. F. Burgio, H.-J. Schulze, and D. Zappal\`a
\\
INFN Sezione di Catania,
Dipartimento di Fisica e Astronomia, Universit\`a di Catania,
Via Santa Sofia 64, 95123 Catania, Italy
}

\date{\today}
\pubyear{2018}

\begin{document}
\label{firstpage}
\pagerange{\pageref{firstpage}--\pageref{lastpage}}
\maketitle

\begin{abstract}
We model the cooling of hybrid neutron stars
combining a microscopic nuclear equation of state in the Brueckner-Hartree-Fock approach
with different quark models.
We then analyze the neutron star cooling curves
predicted by the different models and single out the preferred ones.
We find that
the possibility of neutron p-wave pairing can be excluded in our scenario.
\end{abstract}



\section{Introduction}

The recently observed first neutron star (NS) merger event
\citep{merger,merger2,merger3}
has provided novel strong constraints on the nuclear equation of state (EOS)
by means of information on the tidal deformability and the correlated NS radius
\citep{radice,pascha,drago4,jinbiao}.
This finding has been strengthened by the recent estimates of the mass and radius
of the isolated 205.53~Hz pulsar PSR J0030+0451,
which was observed using the
{\it Neutron Star Interior Composition Explorer} (NICER).
A Bayesian approach to analyze its thermal X-ray waveform yielded
$M=1.44^{+0.15}_{-0.14}\ms$
and
$R=13.02^{+1.24}_{-1.06}\;$km
(68\% credibility interval)
\citep{Miller_2019,Riley_2019},
thus improving the astrophysical constraints on the EOS
of cold, catalyzed matter above nuclear saturation density.
Also the recent update of the currently observed most massive NS,
now MSP J0740+6620 with $M=2.14^{+0.10}_{-0.09}\ms$
(68.3 \% credibility interval) \citep{cromartie}
has further tightened the constraints on the NS EOS.

There is therefore currently a great interest
to single out theoretical EOSs (still) compatible
with the new constraints,
and we have investigated this issue in the framework of the
Brueckner-Hartree-Fock (BHF) approach \citep{drago4,jinbiao,fmic,jinbiao20}.

Another independent source of observational constraints is the
information on the cooling of isolated NSs and of the accreting reheated NSs
in X-ray transients in quiescence (XRTQ)
\citep{yako14,Beznogov1,Beznogov2}.
We have also analyzed this aspect of the BHF EOSs
and found them fully compatible with all current cooling data
\citep{2016MNRAS,2018MNRAS,2019MNRAS}.
In this regard a particular feature of the BHF EOSs is their
fairly large proton fraction in betastable NS matter,
which allows the onset of the very strong direct Urca (DU) cooling process
already in fairly light NSs,
in contrast to the original `standard cooling paradigm' \citep{minimal}.
This type of cooling has therefore to be suppressed
by nucleonic superfluidity over a sufficiently large range of density/NS mass.
We have shown in
\cite{2016MNRAS,2018MNRAS,2019MNRAS}
that such a scenario can be straightforwardly accomplished
assuming standard BCS proton $s$-wave pairing inside the star.
The predictions are robustly compatible with all current cooling data.

The present work is a further extension of this scenario.
We investigate here the effects of an eventual hadron-quark phase transition
in NS matter,
and whether this invalidates the conclusions made so far.
In particular, are there unique signals that would allow to
confirm or reject the presence of such a scenario?
As in nucleonic matter,
the presence of superfluidity in quark matter (QM)
might change the results by orders of magnitude.
However, at this stage of our investigation,
we disregard this possibility,
and study the dependence of the results on the choice of different quark models
without color superfluidity.
Some further comments will be given in the conclusions, however.

There have been several studies in the past
that include quark matter in the cooling process as, for instance,
\cite{iwa80}, where the enhancement of the neutrino emissivity
due to beta decay of quarks was first taken into account.
More recently, the cooling of hybrid stars was studied with quark matter
described by the MIT bag model
(with the parameters of the model adjusted to optimize the fit
to the observational cooling data) in \cite{negre12},
or by the Nambu-Jona-Lasinio model with vector interactions
(and particularly focused  on the cooling of the neutron star in Cassiopeia A)
in \cite{carval15},
or even by a non-local extension of the Nambu-Jona-Lasinio model
in \cite{spin16}.

In addition, the effects of quark superfluidity,
as well as of superconducting quark phases have been studied respectively
in \cite{exo1} and in \cite{Blaschke:1999,gri01,gri05},
and, more recently, in \cite{noda13,sedrak16}.
The role of the quark gaps and of their size has been analyzed.
In particular, in \cite{sedrak16} a transition from a fully gapped,
two-flavor color-superconducting quark phase to a
crystalline or an alternative gapless, color-superconducting quark phase
is considered to explain the Cassiopeia A fast cooling data.

Furthermore, the thermal evolution of strange stars is discussed
in \cite{schaab97} and in \cite{cheng13},
where the specific features of a color-flavor-locked phase are investigated;
the enhanced emissivity due to the combination of Fermi liquid effects
and non-Fermi liquid effects is studied in \cite{schwenzer04},
while the thermodynamic and transport properties of the isotropic
color-spin-locking phase of two-flavor superconducting quark matter
are considered in \cite{berder16}.


This paper is organized as follows.
In Section~\ref{s:eos} we give a brief overview of the theoretical framework,
namely the BHF formalism adopted for the nuclear EOS,
the various QM EOSs,
and the phase transition construction that is used.
Also the various cooling processes
and related nucleonic pairing gaps are reviewed.
Section~\ref{s:res} is devoted to the presentation and discussion
of the results for stellar structure and the cooling diagrams.
Conclusions are drawn in Section~\ref{s:end}.

\section{Formalism}
\label{s:eos}

\subsection{Hadronic Phase}
\subsubsection{Nuclear equation of state}

In our model, we derive the EOS of nuclear matter
within the Brueckner-Bethe-Goldstone many-body theory,
which is based on the resummation of the perturbation expansion of
the ground-state energy \citep{1976Jeu,1999Book,2012Rep}.
The original bare nucleon-nucleon ($NN$) interaction is systematically replaced
by an effective interaction that describes the in-medium scattering processes,
i.e., the so-called $G$-matrix,
that takes into account the effect of the Pauli principle
on the scattered particles and the in-medium single-particle (s.p.)
potential $U(k)$ felt by each nucleon.
The $G$-matrix satisfies the self-consistent equations
($\hbar=c=1$)
\be
  G(\rho,x;E) = V + V \;\text{Re} \sum_{1,2}
 \frac{|12 \rangle (1-n_1)(1-n_2) \langle 1 2|}
 {E - e_1-e_2 +i0} G(\rho,x;E) \:
\label{e:g}
\ee
and
\be
 e_1 = \frac{k_1^2}{2m_1} + U_1 \:,\quad
 U_1(\rho,x) = {\rm Re} \sum_2 n_2
 \langle 1 2| G(\rho,x;e_1+e_2) | 1 2 \rangle_a \:,
\label{e:u}
\ee
where
the multi-indices 1,2 denote in general momentum, isospin, and spin,
$n_1=\theta(k^{(1)}_F-k_1)$ is the momentum distribution,
$x=\rho_p/\rho$ is the proton fraction, and
$\rho_p$ and $\rho$ are the proton and the total baryon density, respectively.
The energy density is obtained as
\be
 \epsilon = \sum_{i=n,p} 2\sum_k n_i(k)
 \left( {k^2\over 2m_i} + {1\over 2}U_i(k) \right) \:.
\ee

Therefore,
in the Brueckner approach the only input required is the bare $NN$ potential $V$
in the Bethe-Goldstone equation~(\ref{e:g}),
for which we use in this paper the Argonne $V_{18}$ \citep{v18} potential
supplemented by a consistent meson-exchange three-nucleon force (TBF),
which allows to reproduce correctly the nuclear matter saturation point
\citep{1989Grange,2002Zuo,Li2008bp,zhl1}
and other properties of nuclear matter around saturation \citep{jinbiao20}.
We remark that the value of the maximum mass $\mmax=2.34\ms$
of the V18 EOS
is larger than the current observational lower limits
\citep{demorest2010,heavy2,fonseca16,cromartie}.
Regarding the radius,
we found in \cite{drago4,jinbiao} that
the value of a 1.4-solar-mass NS, $R_{1.4}=12.33\;$km,
fulfills the constraints derived from
the tidal deformability in the GW170817 merger event.
It is also compatible with estimates of the mass and radius
of the isolated pulsar PSR J0030+0451
observed by NICER \citep{Miller_2019, Riley_2019},
$M=1.44^{+0.15}_{-0.14}\ms$ and
$R=13.02^{+1.24}_{-1.06}\;$km.

We remind that the neutron and proton effective masses
are important ingredients in the cooling simulations,
and they are expressed in terms of the s.p.~energy $e(k)$,
\be
 \frac{m^*(k)}{m} = \frac{k}{m} \left[ \frac{d e(k)}{dk} \right]^{-1} \:.
\ee
In our previous cooling simulations \citep{2016MNRAS},
we used these effective masses derived self-consistently
in the BHF approach \citep{meff}.
However, we found that their effect is small compared to other uncertainties
regarding the cooling,
and therefore in this paper we use the bare nucleon mass for simplicity.

For completeness,
we mention that for the calculation of the stellar structure we use the EOSs
by Feynman-Metropolis-Teller \citep{fey} and Baym-Pethick-Sutherland \citep{BPS}
for the outer and inner crusts, respectively.

\subsubsection{Nuclear cooling processes}
\label{s:cp}

NS cooling is over a vast domain of time
($10^{-10}-10^5\,\text{yr}$)
dominated by neutrino emission due to several microscopic processes
\citep{2001rep,2006ARNPS,2006PaWe,2007LatPra,potrev}.
The theoretical analysis of these reactions requires the knowledge of the
elementary matrix elements,
the relevant beta-stable nuclear EOS, and, very important,
the superfluid properties of the stellar matter, i.e.,
the gaps and critical temperatures in the different pairing channels.

In a non-superfluid NS,
the strongest reactions are the baryon DU processes
(neutron $\beta$ decay and inverse in thermal equilibrium):
\be
 n \ra p + e^-\! + \bar{\nu}_e
\quad \textrm{and} \quad
 p + e^- \ra n + \nu_e \:,
\label{e:DU}
\ee
which are threshold reactions open at densities dependent on the adopted EOS,
due to the energy and momentum conservation \citep{1991LP}.
If they are forbidden,
the main cooling reactions operating in the NS core are the so-called
modified Urca (MU) processes:
\be
 n + N \ra p + N + e^-\! + \bar{\nu}_e
\quad \textrm{and} \quad
 p + N + e^- \ra n + N + \nu_e \:,
\label{e:MU}
\ee
where $N$ is a spectator nucleon that ensures momentum conservation,
and the nucleon-nucleon bremsstrahlung (BS) reactions:
\be
 N+N \ra N+N+ \nu+\bar{\nu} \:,
\ee
with $N$ a nucleon and
$\nu$, $\bar{\nu}$ an (anti)neutrino of any flavor.
Those reactions are abundant and their rate increases with the baryon density,
but they are much less efficient than the DU ones,
thus producing a slow cooling \citep{2001rep}.

For the V18 EOS used in this work,
the DU process sets in at a proton fraction $x_p=0.135$
corresponding to the nucleon density $\rdu=0.37\,$\fm3
of beta-stable and charge-neutral matter,
and a associated NS mass $\mdu=1.01\ms$, as shown in Table~\ref{t:eos}.
Therefore practically all NSs can potentially cool very fast
and slow cooling has to be achieved by superfluid suppression.

We finally remind the possible strong constraints on NS cooling imposed by the
speculated very rapid cooling of the supernova remnant Cas~A
\citep{2009Nat,2010HeiHo,2013Elsha,2015HoPRC},
which we have studied in detail in \cite{2016MNRAS}.
As the observational claims remain highly debated \citep{casno1,casno2},
we do not consider this scenario in this work.

\subsubsection{Nuclear pairing}
\label{s:gaps}

NS cooling is affected not only by the dense matter composition,
but also very strongly by the neutron and proton superfluidity,
as predicted by many microscopic theories \citep{2001rep}.
These superfluids are produced by the $pp$ and $nn$ Cooper pairs formation
due to the attractive part of the $NN$ potential,
and are characterized by a critical temperature
$T_c \approx 0.567\Delta$
related to the pairing gaps in the $^1S_0$ and $^3PF_2$ channels.
The presence of superfluid gaps in the nucleon energy spectrum
reduces the neutrino reaction rates,
and the corresponding neutrino emissivity is exponentially reduced,
together with the specific heat of that component.

On the other hand, the baryon superfluidity initiates a specific
neutrino emission due to Cooper pairing of nucleons,
called the pair breaking and formation (PBF) processes.
These processes take place only in the presence of superfluidity,
with the energy released in the form of a neutrino-antineutrino pair
when a Cooper pair of baryons is formed.
This happens when the temperature reaches $T_c$ of a given type of baryons,
becomes maximally efficient when $T \approx 0.8\,T_c$,
and then is exponentially suppressed for $T \ll T_c$ \citep{2001rep}.

In \cite{2016MNRAS,2018MNRAS,2019MNRAS}
we concluded that a very good description of cooling properties can be obtained
within our BHF approach
by just using the (eventually scaled) BCS values in the p1S0 channel and
disregarding any pairing in the n3P2 channel.
The complete suppression of the 3P2 gaps could be caused by
polarization corrections \citep{pol1,ppol1,ppol2,pol,pol3,ppol3},
which for the 1S0 channel are known to be repulsive,
but for the 3P2 are still essentially unknown;
and this might change the value of these gaps even by orders of magnitude.

For the V18 EOS, the p1S0 pairing gap extends up to a baryon density
$\r1s0=0.60fm^{-3}$ and a corresponding NS mass $\m1s0=1.92\ms$ \citep{2019MNRAS}.
Thus heavier nucleonic stars may exhibit very rapid unquenched DU cooling.
This scenario might change in the case that a hadron-quark phase transition
takes place before $\r1s0$.
In the Gibbs nucleon-quark mixed phase,
which we discuss in this paper,
the nucleonic pairing gaps remain active in the nucleonic component
(we disregard any finite-size effects in the work),
and vanish at a baryon density slighty different from $\r1s0$.
This will be analyzed in detail later.

\subsection{Quark Phase}

\subsubsection{The Dyson-Schwinger model}
\label{s:qm}

One of the models we adopt for cold dense QM
is based on the Dyson-Schwinger (DS) equations of the QCD quark propagator,
described in detail in our previous papers \citep{Chen11,Chen15a,Chen16}.
In this approach,
the quark $q=u,d,s$ momentum distribution function is obtained as
\be
 f_q(|\pv|;\mu) =
 \frac{1}{4\pi} \int_{-\infty}^\infty \! dp_4 \,
 {\rm tr}_{\rm D}\big[-\gamma_4 S_q(p;\mu)\big] \:,
\label{e:nqmuf1}
\ee
and the number density and pressure can be calculated as follows
\be
 n_q(\mu) = 6 \int\frac{d^3\!\pv}{(2\pi)^3} \, f_q(|\pv|;\mu) \:,
\label{e:nqmu}
\ee
\be
 p(\mu_u,\mu_d,\mu_s) = - \bds + \sum_{q=u,d,s}
 \int_{\mu_q^0}^{\mu_q}\! d\mu \,n_q(\mu) \:.
\label{e:pds}
\ee
The fundamental quantity in Eq.~(\ref{e:nqmuf1})
is the quark propagator at finite chemical potential $\mu$.
Within the DS model, it can be written as
\bal
 S(p;\mu)^{-1} &=
 Z_2 \big[ i{\bm \gamma}{\bm p} + i \gamma_4 (p_4+i\mu) + m_q \big]
\\
 &+ Z_1\!\int\!\!\! \frac{d^4q}{(2\pi)^4} g^2(\mu) D_{\rho\sigma}(k;\mu)
 \frac{\lambda^a}{2} \gamma_\rho S(q;\mu) \Gamma^a_\sigma(q,p;\mu) \:,
\label{gensigma}
\eal
where $\lambda^a$ are the Gell-Mann matrices,
$g(\mu)$ is the coupling strength,
$D_{\rho\sigma}(k;\mu)$ the dressed gluon propagator, $k=p-q$,
and $\Gamma^a_\sigma(q,p;\mu)$ the dressed quark-gluon vertex
at finite chemical potential.
In order to solve the equation,
truncations for the quark-gluon vertex and gluon propagator are necessary.
For the vertex we use the bare one, i.e.,
$\Gamma_\sigma^a=\frac{\lambda^a}{2}\gamma_\sigma$.
For the dressed gluon propagator,
we employ the scheme with an infrared-dominant interaction
modified by the quark chemical potential \citep{Chen11,Jiang13},
\be
 g^2(\mu) D_{\rho \sigma}(k,\mu) =
 4\pi^2 d \frac{k^2}{\omega^6} e^{-\frac{k^2+\al\mu^2}{\omega^2}}
 \Big(\delta_{\rho\sigma}-\frac{k_\rho k_\sigma}{k^2}\Big) \:.
\label{gaussiangluonmu}
\ee

The parameters $\omega$ and $d$ in this equation
are discussed in \cite{Alkofer02,Chang09}:
$\omega$ represents the energy scale in nonperturbative QCD,
like $\Lambda_\text{QCD}$,
and $d$ controls the effective coupling strength.
Their values as well as the quark masses are obtained
by fitting light ($\pi$ and $K$) meson properties
and the chiral condensate in vacuum \citep{Alkofer02,Chang09},
and we use the set
$\omega=0.5\;\text{GeV}$ and $d=1\;\text{GeV}^2$.
We choose the quark masses
$m_{u,d}=0$ and $m_s=115\;\text{MeV}$.
In order to model the reduction rate of the effective interaction
with increasing chemical potential,
a phenomenological parameter $\alpha$ was introduced.
Obviously, $\alpha=\infty$ corresponds to a noninteracting system
at finite chemical potential, i.e.,
a simple version of the MIT bag model.
We remark that larger $\alpha$ corresponds to a stiffer quark matter EOS.

Apart from $\alpha$, the bag constant $\bds$ in Eq.~(\ref{e:pds})
is another important parameter in our model.
As discussed in \cite{Chen11,Chen16},
$\bds\approx90$\mfmrm
can be obtained from the vacuum pressure in the massless 2QM case in our model,
but there are ambiguities when including strange quarks.
In this work we treat it as a further phenomenological parameter
like the reduction rate $\alpha$.
We choose the values $\bds=138$\mfmrm and $42$\mfmrm
for $\al=1.5$ and $\al=1.0$, respectively,
in order to fix the maximum mass of the hybrid models to $M=2.1\ms$,
consistent with the current constraint \citep{cromartie}.
This condition imposes a smaller $\bds$ for $\alpha=1.0$.
The effect of smaller $\bds$ makes the EOS of QM stiffer,
but at the same time it lowers the hadron-quark phase transition point
and makes the total EOS of the hybrid star softer.
This results in a reduction of the maximum mass.
We notice that $\bds$ cannot be arbitrarily low,
as one has to ensure that the pressure(energy density) of 2QM should be
lower(larger) than that of symmetric nuclear matter at low density
\citep{Haensel2007,Chen16}.
The DS1.0 model is an extreme choice close to this limit
in order to enforce a QM onset density $\rqm$ as low as possible.

\subsubsection{The field correlator model}

In addition to the DS model,
we also consider the field correlator model (FCM)
to represent the dense QM phase,
as the latter has already been used in the description of hybrid stars
with the aim of determining the mass-radius relation of the star,
and deducing possible constraints on the parameters of the model
\citep{Baldo:2008en,Bombaci:2012rv,Plumari:2013ira,Alford:2015dpa,Burgio:2015zka}.
Within this approach, discussed in detail in \cite{DiGiacomo:2000irz},
one performs a description of the strong interaction dynamics
in terms of Gaussian correlators of color-electric and color-magnetic fields,
which naturally encompasses the confinement mechanism.

The analysis of the high-density matter in the NS core
requires the extension of the FCM to finite density (and temperature),
which was derived in
\cite{Simonov:2007xc,Simonov:2007jb,Nefediev:2009kn}
in the single-line approximation,
where the leading contribution is given by the interaction
of single quark and gluon lines with the vacuum.
The resulting pressure of the QM phase,
\be
 p_\text{QM} = p_V + p_g + \sum_{q=u,d,s}\!p_q \:,
\label{pqgp1}
\ee
is the sum of the vacuum $p_V$, gluon $p_g$, and quark $p_q$ contributions,
which are reported below.
\be
 p_V= - \frac{(11-2N_f/3)}{32} \frac{G_2}{2}
\label{pqgp2}
\ee
represents the pressure difference between the vacua
in the deconfined and confined phases,
and includes the lattice indication that the gluon condensate $G_2$
is sharply reduced by half at the transition
observed at the critical temperature \citep{DElia:1997sdk,DElia:2002hkf}.
\be
 p_g = \frac{8 T^4 }{3 \pi^2} \int_0^\infty  d\chi \chi^3
 \frac{1}{\exp{(\chi + 9V_1/8T)} - 1} \:
\label{pglue}
\ee
and
\be
 p_q = \frac{T^4}{\pi^2}  \left[
 \phi_\nu \Big( \frac{\mu_q - V_1/2}{T}\Big) +
 \phi_\nu \Big(-\frac{\mu_q + V_1/2}{T}\Big) \right]
\label{pquark}
\ee
are the gluon and quark pressure,
where $p_q$ is intended for each single flavor with mass
$m_q$ ($\nu=m_q/T$) and chemical potential $\mu_q$,
and with
\be
 \phi_\nu(a) = \int_0^\infty du \frac{u^4}{\sqrt{u^2+\nu^2}}
 \frac{1}{\exp{(\sqrt{u^2 + \nu^2} - a)} + 1} \:.
\label{distrib}
\ee

Apart from the external parameters $\mu_q$ and $T$
(and the quark masses $m_q$),
$p_\text{QM}$ depends on two parameters that are peculiar to the model,
namely $V_1$ and $G_2$.
The parameter $V_1$ at zero temperature and density is expressed
in terms of an integral of a fundamental QCD correlator
\citep{DiGiacomo:2000irz},
and indicates the large-distance static particle-antiparticle potential,
but it is not directly measured.
$G_2$ is the gluon condensate,
whose estimate is known from QCD sum rules,
$G_2 \approx 0.012\;\rm{GeV^4}$,
but with an uncertainty of about $50\%$.
In addition, from Eq.~(\ref{pqgp2}) we notice that $G_2$ has the same role
as the bag constant of the MIT bag model and thus,
if one sets $V_1=0$,
the quark pressure $p_q$ becomes the pressure of free quarks and the model
reduces to the simplest version of the bag model.
Therefore $V_1$ represents the main correction
to the free quarks dynamics inside the bag.

Although there exist some speculations on the temperature dependence of
$V_1$ and $G_2$ and on their estimates at the critical temperature
\citep{Simonov:2007xc,Simonov:2007jb,Bombaci:2012rv,Plumari:2013ira},
not very much is known about their dependence on the baryonic density,
which is certainly relevant for the description of the inner core of NSs.
Therefore, for our purpose,
it is preferable to avoid specific assumptions about
the temperature and density dependence
and to treat $V_1$ and $G_2$ as free parameters.
In this spirit, some indications on the phenomenologically acceptable ranges
of $V_1$ and $G_2$,
that predict maximum hybrid star masses compatible with the observational limits,
have been obtained in \cite{Alford:2015dpa,Burgio:2015zka},
suggesting large values of the interaction strength,
$V_1 \sim 100 - 200 \;\text{MeV}$
and rather small values of the gluon condensate,
$G_2 \sim 0.002 - 0.006 \;\text{GeV}^4$.
In the calculations shown in this paper,
we have used $V_1 = 142 \;\text{MeV}$
and $G_2 = 0.006 \;\text{GeV}^4$,
fixing the hybrid star maximum mass to $2.1\ms$,
as for the DS model.

\subsubsection{Gibbs phase transition construction}

We assume that the hadron-quark phase transition is of first order,
and perform the Gibbs construction \citep{Glen1992},
thus imposing that nuclear matter and QM are betastable
and globally charge neutral.
This is at variance with the Maxwell construction,
where the two phases must be separately charge neutral.
However, in our approach the Maxwell construction usually produces unstable
stellar configurations at the onset of the QM phase, see \cite{Chen11}.
We also remind that Maxwell and Gibbs constructions are the extremes
of the generalized Gibbs construction taking into account finite-size
effects \citep{Glen1992,2000Glen,Maru2007,Chen13},
which are however currently quantitatively unknown due to missing
input information.
Therefore neither pure Maxwell nor pure Gibbs constructions are expected
to be realized in nature.

In the purely nucleonic phase,
which consists of baryons ($n,p$) and leptons ($e,\mu$),
the conditions of  beta stability and charge neutrality can be expressed as
\bal
 & \mu_n-\mu_p = \mu_e = \mu_\mu \:,
\\
 & \rho_p = \rho_e + \rho_\mu \:,
\eal
where $\mu_i$ are the chemical potentials
and $\rho_i$ the particle number densities.
Similarly the pure QM phase,
which contains three-flavor quarks ($u,d,s$) and leptons ($e,\mu$),
should satisfy the constraints
\bal
 & \mu_u+\mu_e = \mu_u+\mu_\mu = \mu_d = \mu_s \:,
\\
 & {2\rho_u-\rho_d-\rho_s\over 3} - \rho_e - \rho_\mu = 0 \:.
\eal

In the mixed phase according to the Gibbs construction,
the hadron and quark phases coexist
in thermodynamic equilibrium with each other \citep{gle}.
This can be expressed as\be
 \mu_i = b_i \mu_B - q_i \mu_e \ , \quad p_H = p_Q = p_M \:,
\ee
where $b_i$ and $q_i$ denote baryon number and charge of the particle species
$i=n,p,u,d,s,e,\mu$ in the mixed phase.
Those equations are solved together with
the global charge neutrality condition
\be
 \chi\rho_c^Q + (1-\chi)\rho_c^H = 0 \:,
\ee
where $\rho_c^Q$ and $\rho_c^H$ are the charge densities
of quark and nuclear matter,
and $\chi$ is the volume fraction occupied by QM in the mixed phase.
The baryon density $\rho_M$ and the energy density $\eps_M$
of the mixed phase are then
\bal
 & \rho_M = \chi\rho_Q + (1-\chi)\rho_H \:,
\\
 & \eps_M = \chi\eps_Q + (1-\chi)\eps_H \:,
\eal
where $\rho_Q$ and $\rho_H$ ($\eps_Q$ and $\eps_H$) are the baryon (energy) densities
of quark and nuclear matter in the mixed phase.

\subsubsection{Quark matter cooling}

Similar to nuclear matter,
the most powerful neutrino emission from QM is also the DU process
via the direct and inverse $\beta$ decay
\citep{iwa82,2001rep,yako14},
\bal
 & d \ra u + e^-\! + \bar{\nu}_e
 \qquad \textrm{and} \qquad
 u + e^- \ra d + \nu_e \:,
\\
 & s \ra u + e^- + \bar{\nu}_e
 \qquad \textrm{and} \qquad
 u + e^- \ra s + \nu_e \:.
\label{e:DUQ}
\eal
In contrast to the nucleonic DU process
which requires a threshold density \citep{1991LP},
the quark DU process switches on immediately at the onset of QM
since the electron fraction is high enough.
In fact, if the density is large, the quark DU reactions could be
completely switched off because of the too small electron fraction
not fulfilling the Fermi momentum conservation \citep{Duncan}.

The main cooling ingredient is thus the quark DU emissivity,
which for unpaired QM has been considered by \cite{iwa80,iwa82},
where $u,d$ quarks were considered as massless.
The emission rates are
\bal
 & \epsilon_Q^{(d)} = \frac{914}{315} \alpha_s (G_F\cos\theta_c)^2
 k_F^{(d)} k_F^{(u)} k_F^{(e)} T^6 \:,
\\
 & \epsilon_Q^{(s)} = \frac{457\pi}{840} (1-\cos\theta_{34}) (G_F\sin\theta_c)^2
 \mu_s k_F^{(u)} k_F^{(e)} T^6 \:,
\label{e:eDUQ}
\eal
where $\alpha_s$ is the strong coupling constant,
$G_F$ is weak coupling constant,
$\theta_c$ is the Cabibbo angle
($\cos\theta_c^2 \approx 0.948$),
$\theta_{34}$ comes from the decay kinematics,
$k_F^{(i)}(i=u,d,e)$ is the Fermi momentum of each particle,
and $T$ is the temperature.

\begin{figure}
\vspace{-0mm}
\centerline{\includegraphics[scale=0.35]{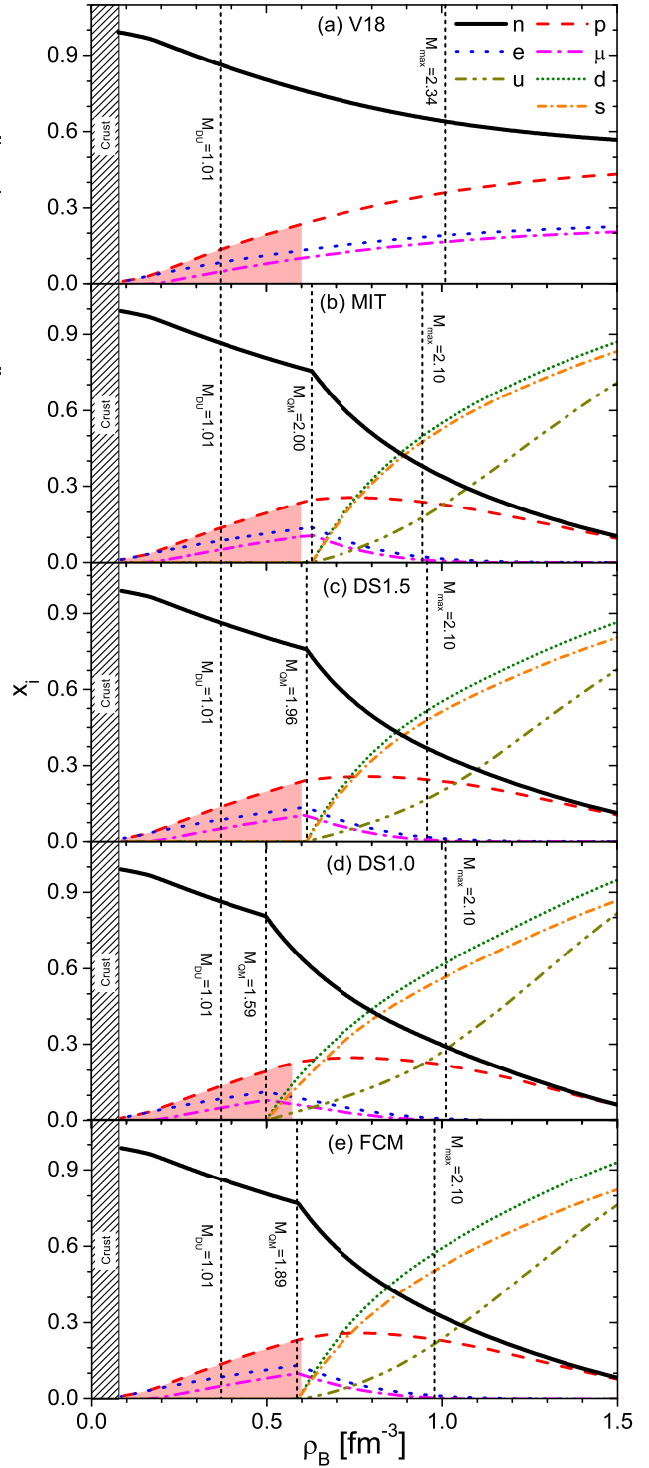}}
\vspace{-0mm}
\caption{
Composition of betastable stellar matter with the Gibbs construction
for different quark models.
Vertical lines indicate
the onset of nucleonic DU cooling,
the onset of the QM mixed phase,
and the maximum central density.
Corresponding mass values of NSs with those central densities are given.
The shaded areas indicate the range over which the p1S0 gap is active.
}
\label{f:xi}
\end{figure}

The absence of the strong coupling constant in Eq.~(\ref{e:eDUQ})
indicates that the $\beta$ decay
of the $d$ quark is due to the strong interaction,
whereas the one of the $s$ quark is triggered by the finite quark mass
\citep{iwa82}.
This results in somewhat different neutrino emission rates.
In general, the $s$ quark DU gives a smaller emissivity \citep{iwa82,2001rep},
but above a certain density it could be the most prominent emission.
As can be seen in Fig.~\ref{f:xi},
the fraction (or particle number density) of $s$ quarks is always smaller
than the one of the $d$ quarks,
meaning that in the $us$ branch the Fermi momentum conservation
is easier to achieve than in the $ud$ branch.
It might happen that the $s$ quark DU is active,
and the $d$ quark DU is inactive at high density,
as we found in the DS model with $\alpha=1.5$, for example.

Once the density keeps increasing,
the $s$ quark DU will also be switched off.
In this case, the quark MU and BS reactions
would be the dominant neutrino processes in QM \citep{gri01}.
However, this requires that the density reaches values of
$\rho_B(\text{MIT, DS1.5, DS1.0, FCM}) = (3.83, 1.93, 1.65, 7.79)\,$\fm3,
which are much larger than the central density of the most massive star,
as shown in Table~\ref{t:eos}.
On the other hand,
in our models
the quark DU processes are always active inside the star's core.

\begin{table}
\caption{
Characteristic properties of several EOSs:
densities $\rho$ (in \fm3) and
corresponding NS masses $M$ (in $\ms$) with that central density
characterising the DU cooling onset, QM onset, the vanishing of the p1S0 gap,
and maximum-mass configuration.
}
\def\myc#1{\multicolumn{1}{c}{$#1$}}
\renewcommand{\arraystretch}{1.2}
\begin{tabular}{lccccccccc}
  \hline\hline
  EOS & $\rdu$ & $\mdu$ & $\rqm$ & $\mqm$ & $\r1s0$ & $\m1s0$ &
  $\rho_\text{max}$ & $\mmax$ \\
  \hline
  V18   & 0.37 & 1.01  &   -   &   -   & 0.599 & 1.92   & 1.010 & 2.34\\
  MIT   & 0.37 & 1.01  & 0.629 & 2.00  & 0.599 & 1.92   & 0.944 & 2.10\\
  DS1.5 & 0.37 & 1.01  & 0.614 & 1.96  & 0.599 & 1.92   & 0.960 & 2.10\\
  DS1.0 & 0.37 & 1.01  & 0.498 & 1.59  & 0.576 & 1.80   & 1.009 & 2.10\\
  FCM   & 0.37 & 1.01  & 0.588 & 1.89  & 0.599 & 1.91   & 0.977 & 2.10\\
  \hline \hline
\end{tabular}
\label{t:eos}
\end{table}

\begin{figure}
\vspace{-6mm}
\centerline{\hspace{0mm}\includegraphics[scale=0.35]{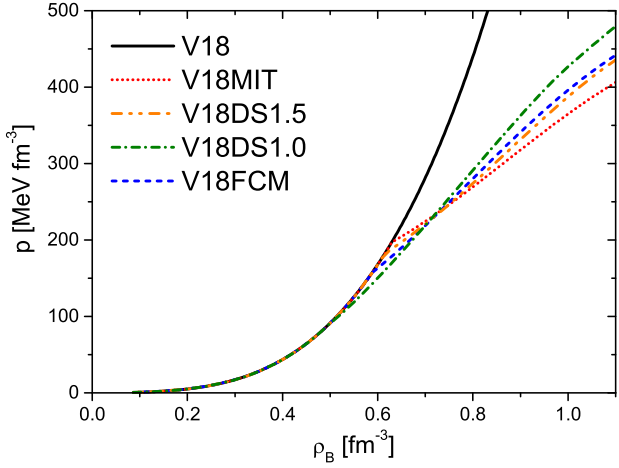}}
\vspace{-4mm}
\caption{
The EOS for the different models.
The solid black curve represents the purely hadronic case,
whereas the broken colored curves are the hybrid EOSs.
}
\label{f:pn}
\end{figure}

\begin{figure}
\vspace{-5mm}
\centerline{\hspace{-5mm}\includegraphics[scale=0.34]{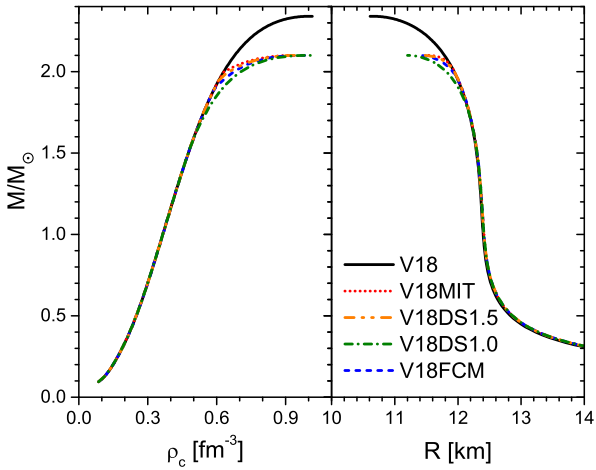}}
\vspace{-3mm}
\caption{
Gravitational mass vs.~central density (left panel) and radius (right panel)
for the different models.
The maximum mass for all hybrid EOSs is, by construction, $\mmax=2.1\ms$.
}
\label{f:mr}
\end{figure}

\begin{figure}
\vspace{-5mm}
\centerline{\includegraphics[angle=0,scale=0.35]{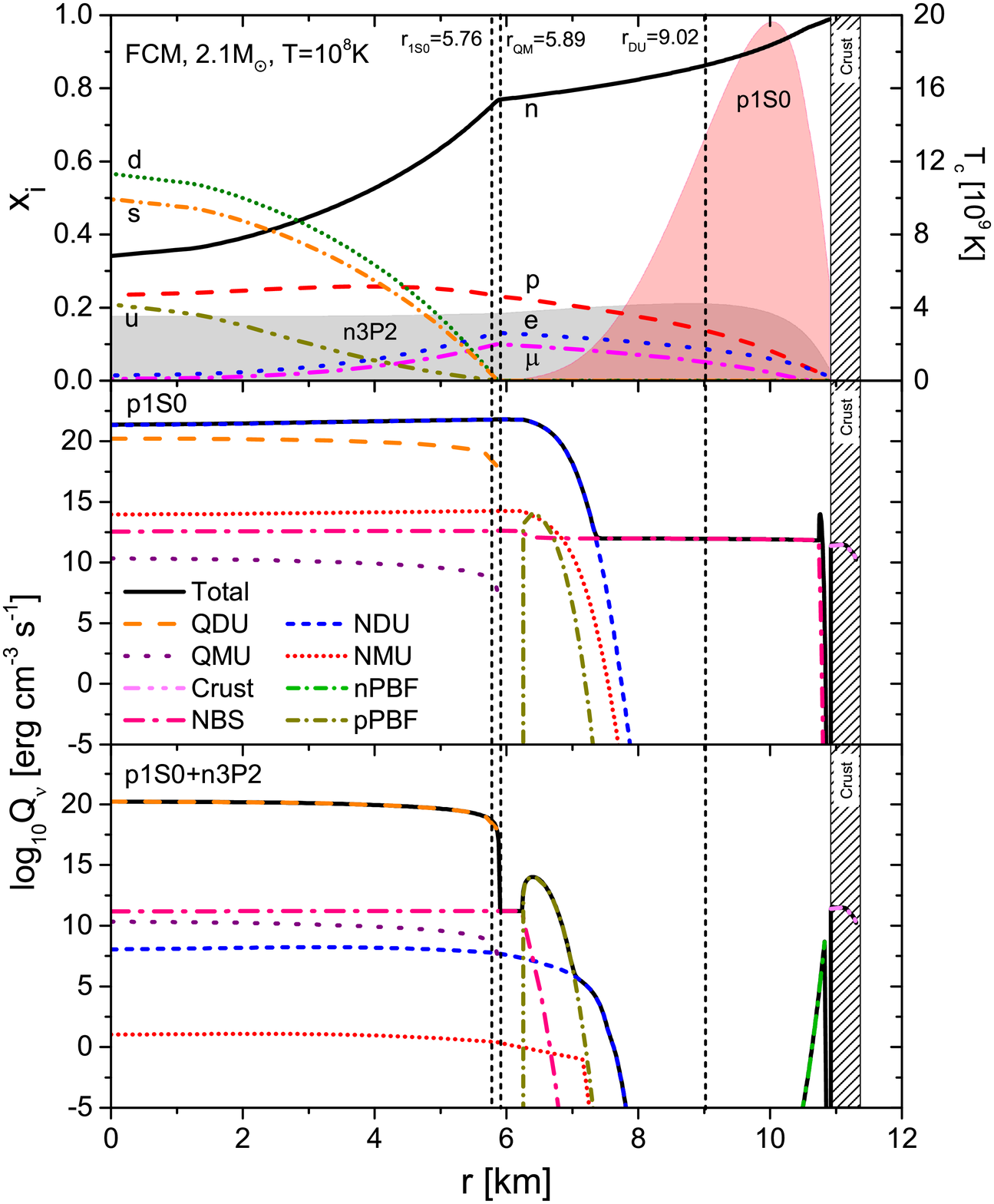}}
\vspace{-41mm}
\caption{
Composition and pairing gaps (upper panel)
and neutrino emissivities at $T=10^8$K (lower panels)
as a function of the radial distance $r$
for a hybrid star with mass $M=2.1\ms$ and the FCM EOS.
}
\label{f:pro}
\end{figure}

\begin{figure*}
\vspace{-2mm}
\centerline{\includegraphics[angle=0,scale=0.35]{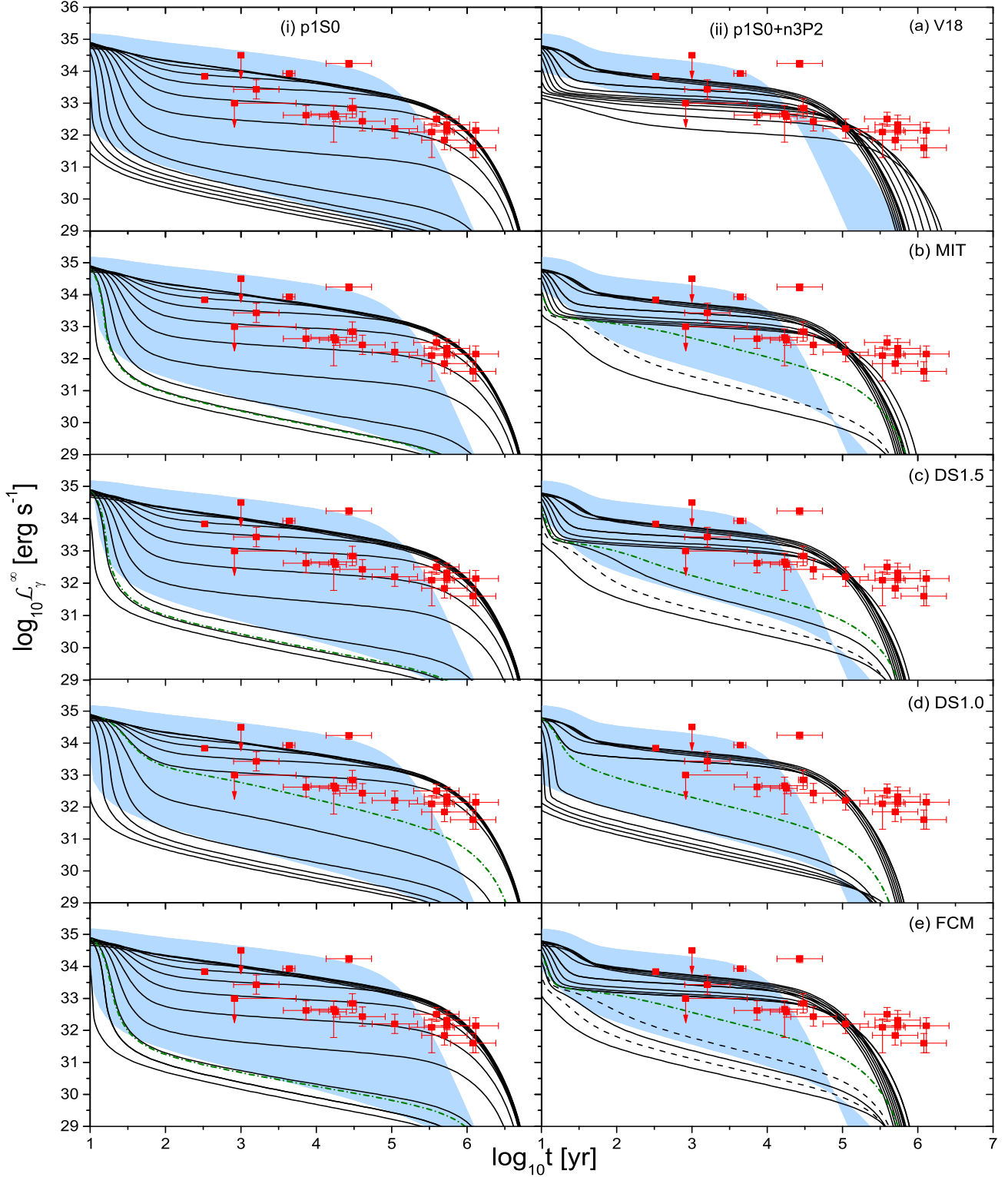}}
\vspace{-8mm}
\caption{
Cooling curves with/out n3P2 pairing,
for different NS masses $M/\ms=1.0,1.1,\ldots,2.1$
(decreasing solid black curves).
Eventual black dashed curves indicate the $M/\ms=1.95,2.05$ results.
The dash-dotted green curves mark the NS mass $\mqm+0.02\ms$
at which the QM DU process has just set in.
The black curves are obtained with a Fe atmosphere and the shaded areas
cover the same results obtained with a light-elements
($\eta=10^{-7}$) atmosphere.
The data points are from \protect\cite{Beznogov1}.
See text for more details.}
\label{f:lt}
\end{figure*}

\begin{figure*}
\vspace{-4mm}
\centerline{\includegraphics[angle=0,scale=0.62]{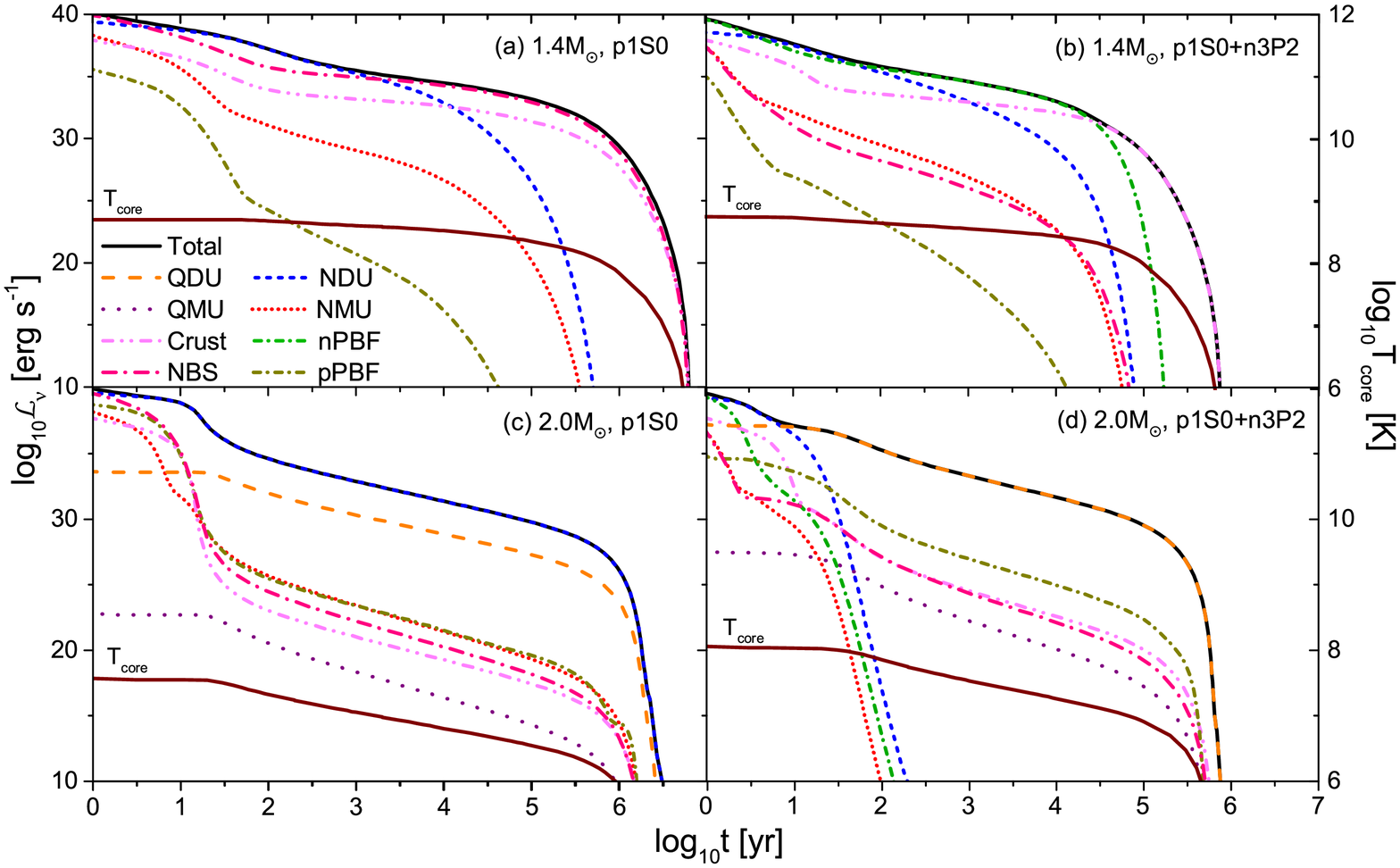}}
\vspace{-1mm}
\caption{
Contributions of the various cooling processes to the total luminosity
as a function of time
for $M=1.4\ms$ and $2.0\ms$ NSs with the FCM EOS.
Results with/out effects of the n3P2 gaps are compared.
The core temperature is also shown (rhs scale).
}
\label{f:em}
\end{figure*}

\section{Results and discussion}
\label{s:res}

We now present the numerical results comparing the hadronic V18 EOS and
the various hybrid models.
The parameters of all quark models have been adjusted in
such a way as to decrease the maximum mass from the value $\mmax=2.34\ms$
of the purely hadronic V18 EOS down to the same value $\mmax=2.10\ms$
in all hybrid star cases.
This requires fairly large QM threshold densities $\rqm$
and associated masses $\mqm$,
see Table~\ref{t:eos}.
The biggest QM content is achieved by the DS1.0 model with
$\rqm\approx0.5\;$\fm3 and $\mqm\approx1.6\ms$.
Lowering further the values of $\rqm$ leads to too low maximum masses
of the hybrid models.
In the same table we also list the density $\r1s0$
at which the p1S0 gap vanishes
and the corresponding gravitational mass,
which will be important for the forthcoming discussion.
For completeness, the values of the central densities
of the maximum-mass configuration are also reported.

\subsection{Composition and structure}
\label{s:mr}

In Fig.~\ref{f:xi} we compare the particle fractions of all models.
In the upper panel results are displayed for the purely nucleonic case,
whereas in the other panels the populations of the hybrid models are plotted.
The vertical dashed lines represent the values of the baryon density
at which the nucleonic DU process starts and the corresponding mass $\mdu$,
the values of the QM onset for the mixed phase and its mass $\mqm$,
and finally the central density of the maximum-mass configuration $\mmax$.
We notice that in all cases the onset of the DU process takes place
at a smaller density than the one of the mixed phase,
which depends on the adopted model for the QM phase.
In all cases, the maximum-mass configurations contain still
more than $50\%$ of nucleonic matter in their center,
while pure QM is only reached at extreme densities,
not present in hybrid stars.
The DS1.0 model features the biggest QM content,
and in this case the p1S0 gap extends into the mixed phase.
A similar behavior is slightly present also in the FCM,
whereas in the MIT and DS1.5 models
the p1S0 gap is active only in the pure nucleonic phase.

Once the composition is known,
the resulting EOSs $p(\rho)$ can be calculated.
They are shown in Fig.~\ref{f:pn}.
One notes again the lowest onset density of the mixed phase for the DS1.0 EOS,
see also Table~\ref{t:eos} and Fig.~\ref{f:xi},
and a strong softening due to the presence of QM.
The consequence is the decrease of the NS maximum mass,
as shown in Fig.~\ref{f:mr} for the different models.
We remind the reader that, by construction,
$\mmax=2.1\ms$ for all hybrid models.

\subsection{Cooling reactions}
\label{s:rescool}

The NS cooling simulations are carried out using the
widely known code {\tt NSCool} \citep{Pageweb},
which comprises all relevant cooling reactions:
nucleonic DU, MU, PBF, and BS,
including modifications due to p1S0 and n3P2 pairing,
as well as QM DU, MU, and BS.

The role played by the different processes is illustrated in detail
in Fig.~\ref{f:pro}
for a V18+FCM hybrid star with (maximum) mass $M=2.1\ms$.
The upper panel displays the nucleonic, leptonic, and quark populations
(curves)
together with the p1S0 and n3P2 critical temperatures
(shaded areas)
vs.~the radial distance,
whereas the lower panels show the various neutrino emissivities
at a temperature $T=10^8\,$K
(corresponding roughly to an age of 10y or 0.1y with/out n3P2 pairing)
for the different cooling channels.
The central panel employs only 1S0 pairing,
whereas the lower panel includes also the n3P2 gap.

We observe that in the mixed phase in the core
(containing up to about $50\%$ QM),
the main contribution to the cooling comes from the QM DU (QDU),
and in the case without n3P2 gap
(central panel)
also from the nucleonic DU (NDU) reaction.
All other reactions are weaker by several orders of magnitude.

At about $r\approx6\;$km QM vanishes
and the cooling is regulated by the nucleonic processes only:
the dominant NDU is active up to $r\approx9\;$km,
where the proton fraction becomes too small.
For this model, the p1S0 gap happens to vanish close to the onset of the
mixed phase,
and therefore NDU cooling is undamped inside the mixed phase
(central panel)
{\em unless} the n3P2 gap is present (lower panel).
The NMU and QMU reactions play only minor roles,
together with NBS,
which is the only relevant cooling process in the outer part of the star
when only p1S0 pairing is active (central panel).

The PBF processes merit a separate discussion:
Due to their nature, they only provide significant and even dominant
contributions when the local critical temperature
(either p1S0 or n3P2)
is slightly above the value of the actual temperature.
Under the conditions chosen for Fig.~\ref{f:pro}
(see the values of $T_c$ in the upper panel),
this occurs around $r\approx6-7\,$km for the p1S0 channel,
and $r\approx10-11\,$km for the n3P2 gap when present.
In the latter case,
due to the concurrent suppression of the NDU process,
the PBF reactions become the most efficient nucleonic cooling process,
which has important consequences for the final luminosity vs.~age plots.

We stress that emissivity plots like Fig.~\ref{f:pro}
depend decisively on the matter composition, i.e., the NS mass,
and the value of the temperature, related to the NS age.
The values $M=2.1\ms$ and $T=10^8\,$K chosen for Fig.~\ref{f:pro}
provide only one particular snapshot of the cooling history.
In particular the zones and magnitudes of PBF cooling inside the star depend
extremely sensitively on the local temperatures,
and the complete cooling history has to be integrated in order
to make quantitative statements,
which we investigate now.

\subsection{Cooling diagrams}

Fig.~\ref{f:lt} shows
the resulting final cooling diagrams for the different models,
namely the luminosity vs.~age is plotted for several NS masses
in the range $1.0,1.1,...,2.1\ms$ (solid black curves).
Eventual dashed black curves represent $M=1.95,2.05\ms$ for a better resolution.
The dash-dotted green curves mark the NS mass $\mqm+0.02\ms$
at which the QM DU process has just set in.
Results employing only 1S0 pairing (left column)
and with n3P2 pairing included (right column)
are compared for the different models.

Our set of observational cooling data comprises the
(age, luminosity) information
of the 19 isolated NS sources listed in \cite{Beznogov1,beloin18},
where it was also pointed out that in many cases the distance to the object,
the composition of its atmosphere, thus its luminosity
and age are rather estimated than measured.
Thus in these cases, we use large ad-hoc error bars (a factor 0.5 and 2)
to reflect this uncertainty.
In order to assess the influence of the heat-blanketing effect
of the atmosphere \citep{potek},
we compare results obtained with
a non-accreted heavy-elements (Fe) atmosphere 
(curves in Fig.~\ref{f:lt})
and one containing also a maximum fraction
$\eta = {g_s}_{14}^2\,\Delta M/M = 10^{-7}$
of light elements from accretion
(shaded areas in Fig.~\ref{f:lt}).

One observes the following general features:
Since $\mdu=1.01\ms$, practically all cooling curves involve
nucleonic DU cooling,
which is however quenched by the p1S0 pairing active up to
$\m1s0\approx1.9\ms$
($\approx1.8\ms$ for the DS1.0 model),
nearly coincident with the onset of the QM phase
and the related rapid QM DU cooling.
Since the QM onset density is fairly large for all quark models,
only high-mass NSs, $M\gtrsim1.9\ms$ ($\gtrsim1.6\ms$ for the DS1.0),
exhibit different cooling behavior for the hybrid models.
No observational data exist currently for such heavy and faint objects
(lying below the dash-dotted green curves in Fig.~\ref{f:lt}).
On the other hand,
very reasonable NS mass distributions can be deduced
when confronting the theoretical curves
with the available cooling data in the figure,
see \cite{2019MNRAS}.

In the right column of Fig.~\ref{f:lt},
we display the cooling curves for the case with the n3P2 gap included.
As already found in \cite{2018MNRAS},
an important conclusion can be drawn regarding the nucleonic pairing:
While very satisfactory results can be obtained
employing only 1S0 pairing,
the addition of n3P2 superfluidity
leads to too fast cooling for all models considered,
such that old and warm NSs cannot be reproduced by any model.
Thus, as for the purely nucleonic EOSs before \citep{2019MNRAS},
we can exclude the possibility of n3P2 pairing in our approach,
even allowing the existence of a phase transition to QM.
This is because only very massive stars are affected by this feature.

In order to better understand the too fast cooling provided by the n3P2 gap,
we show in Fig.~\ref{f:em} the decomposition of the total luminosity
into its various contributions
for `normal' ($M=1.4\ms$, upper panels)
and heavy ($M=2.0\ms$, lower panels) hybrid NSs with the FCM.
The core temperature is also displayed for better understanding.
The main observations are:\\
- For normal stars (no QM),
the eventual n3P2 PBF reaction is the dominant cooling process in panel (b)
due to the fact that the core temperature remains of the order of the
n3P2 critical temperature for most of the cooling history.
This leads to too cold old NSs compared to the available data
(see Fig.~\ref{f:lt}).
The NDU process is completely blocked by the p1S0 gap
that extends throughout the whole star in this case,
and even more by the eventual additional n3P2 gap in (b).
This keeps the core temperature sufficiently high to obtain
warm old NS in agreement with the data in the first panel (a).
The additional n3P2 gap also supresses the NBS (nn) reaction
in (b) compared to (a),
see also Fig.~\ref{f:pro}.\\
- For heavy NSs, the NDU process is unblocked in the mixed phase
when only p1S0 pairing is present in (c),
but becomes completely blocked by the additional n3P2 gap in (d),
see also Fig.~\ref{f:pro}.
The strong QDU reaction is also active in both cases
and dominant in (d) at any time.
Therefore the PBF reactions are not decisive here
and cannot compensate for the blocking of the NDU by the n3P2 gap.
All this leads to much lower core temperatures compared to (a,b),
but warmer stars when including the n3P2 gap and associated blocking of NDU (d)
than not (c).

\section{Conclusions}
\label{s:end}

We studied the cooling of hybrid NSs,
combining a realistic microscopic BHF model for the nucleonic EOS
with different quark models,
joined by a Gibbs phase transition.
The large maximum mass of the nucleonic model is lowered
to a common value of $2.1\ms$ in all hybrid scenarios,
which could be close to the `true' value.

The nucleonic DU cooling process is active in all stars,
but blocked by BCS p1S0 pairing up to the onset of the QM phase.
Therefore only very heavy hybrid stars,
typically $M\gtrsim1.9\ms$,
exhibit rapid QM DU cooling,
while reasonable smooth NS mass distributions
in agreement with current data
are predicted by the effect of nucleonic cooling solely.

An important conclusion can be drawn regarding n3P2 superfluidity:
In all possible scenarios with and without QM,
its presence leads to too rapid cooling of all NSs,
such that the high luminosity of all currently observed old
($t\gtrsim10^5$y) stars cannot be reproduced.
This seems to be a robust result of all models involving nucleonic DU cooling.
This conclusion is also very unlikely to be changed by the effects
of QM pairing that was disregarded in this work.

We have only studied a very limited set of QM EOSs here,
but in general it seems difficult to reconcile an early onset of QM
with a sufficiently small reduction of the maximum mass of the nucleonic EOS.
This was confirmed by the extreme DS1.0 model we studied.
Therefore only heavy NSs could be hybrid stars, and
we thus expect our results to be robust wrt changes of the quark model.

The combined and consistent analysis of different aspects of NS physics
like mergers, radius measurements, and cooling
will allow in the future a more and more accurate derivation of the nuclear EOS
and its constraints.
Regarding the cooling data,
any information on very faint objects of any age
would be most valuable for theoretical progress.

\section*{Acknowledgments}

We acknowledge helpful communication with H.~Chen, D.~Page,
and financial support from ``PHAROS'' COST Action CA16214,
and the China Scholarship Council (No.~201706410092).

\bibliographystyle{mnras}     
\bibliography{coolqm}         
\bsp                          

\label{lastpage}
\end{document}